\documentclass[12pt]{article}
\usepackage{amsmath}
\def\a{\alpha}
\def\b{\beta}

\def\g{\gamma}

\def\t{\theta}

\def\s{\sigma}

\def\be{\begin{equation}}
\def\ee{\end{equation}}
\def\arr{\begin{array}{rll}}
\def\ea{\end{array}}
\def\bea{\begin{eqnarray}}
\def\eea{\end{eqnarray}}
\def\N2{$N{=}2$}

\def\>{\rangle}
\def\<{\langle}
\def\+{\dagger}
\def\={\ =\ }

\begin{document}
%\large
\renewcommand{\thefootnote}{\fnsymbol{footnote}}
\begin{titlepage}
\setcounter{page}{0}
\begin{flushright}
LMP-TPU--11/12  \\
\end{flushright}
\vskip 1cm
\begin{center}
{\LARGE\bf  Near horizon black holes in diverse }\\
\vskip 0.5cm
{\LARGE\bf  dimensions  and integrable models}\\
\vskip 2cm
$
\textrm{\Large Anton Galajinsky \ }
$
\vskip 0.7cm
{\it
Laboratory of Mathematical Physics, Tomsk Polytechnic University, \\
634050 Tomsk, Lenin Ave. 30, Russian Federation} \\
{E-mail: galajin@tpu.ru}

\end{center}
\vskip 1cm
\begin{abstract} \noindent
The near horizon geometry of extremal rotating black hole in arbitrary dimension possesses
$SO(2,1) \times U(n)$ symmetry in the special case that all $n$ rotation parameters are equal.
We consider a conformal particle associated with such a maximally symmetric configuration and derive from it a new
integrable Hamiltonian mechanics with $U(n)$ symmetry. A further reduction of the model is discussed, which is obtained by discarding cyclic variables. A variant of the Higgs oscillator and the P\"oschl--Teller system show up in four and five dimensions, respectively.
\end{abstract}

\vspace{0.5cm}

PACS: 04.70.Bw; 11.30.-j; 02.30.Ik \\ \indent
Keywords: black holes, conformal mechanics, integrable models
\end{titlepage}

\renewcommand{\thefootnote}{\arabic{footnote}}
\setcounter{footnote}0

\noindent
{\bf 1.  Introduction}\\

In recent years there has been considerable interest in various aspects of the Kerr/CFT correspondence \cite{str}.\footnote{By now there is a rather extensive literature on the subject. For recent reviews and further references to the original literature see e.g. \cite{str1,com}.} To a large extend the original proposal in
\cite{str} was motivated by the earlier work \cite{bh}, where it was demonstrated that in the near horizon limit the isometry group $U(1)\times U(1)$ of the extremal Kerr black hole in four dimensions was enlarged to $SO(2,1)\times U(1)$. Note that the first factor is the conformal group in one dimension.

In the conventional formulation of the Kerr/CFT correspondence \cite{str}, one considers excitations around the near horizon extremal Kerr black hole, which are controlled by specific boundary conditions. For every set of boundary conditions there is an associated asymptotic symmetry group formed by diffeomorphisms consistent with the conditions. A key ingredient of the construction is a computation of conserved charges related to the asymptotic symmetry transformations. An asymptotic symmetry is considered to be trivial if the corresponding charge vanishes.
A curios fact about the Kerr/CFT correspondence is that the $U(1)$ factor in the isometry group of the Kerr metric is enhanced to a chiral Virasoro algebra,  while $SO(2,1)$ becomes trivial.

As the $SO(2,1)$ symmetry of the near horizon extremal Kerr solution appears to play a trivial role within the Kerr/CFT correspondence, it is worth studying its implication in a different context. In this work we employ this conformal symmetry to construct new (super)integrable models.

As is well known, each Killing vector field characterizing a background geometry yields an integral of motion of the geodesic equations. In particular, a massive relativistic particle moving near the horizon of an extremal rotating black hole in arbitrary dimension is conformal invariant.\footnote{For recent studies of conformal mechanics related to the near horizon geometries see e.g. \cite{cdkktp}--\cite{sag}.} Recently it was demonstrated in \cite{ar,ar1} that the Casimir element of $so(2,1)$ algebra underlying a generic conformal mechanics gives rise to a
reduced Hamiltonian system called a spherical mechanics. In this work we initiate a systematic study of a spherical mechanics associated with an extremal rotating black hole in arbitrary dimension and derive its Hamiltonian.

In arbitrary dimension, a black hole may rotate in various orthogonal spatial two--planes. Given $n$ independent rotation parameters,
the isometry group of the metric involves ${U(1)}^n$ factor. If all the rotation parameters be equal, the isometry group is enhanced to $U(n)$. In this work we focus on such a maximally symmetric configuration and construct new (super)integrable models, which inherit $U(n)$ symmetry of the background.

In the next section we discuss the near horizon limit of the extremal Kerr black hole in four dimensions. Conformal mechanics associated to it is constructed and a reduced two--dimensional integrable spherical mechanics is derived. A further reduction of the latter to a one--dimensional system, which is obtained  by discarding cyclic variables, is shown to yield a variant of the Higgs oscillator \cite{Higgs}. In Section 3 we consider the near horizon limit of the five--dimensional Myers--Perry black hole. The corresponding conformal mechanics is described and a three--dimensional reduced integrable spherical mechanics is constructed. If the rotation parameters of the black hole be equal to each other, the resulting model is shown to be minimally superintegrable. A further reduction to a one--dimensional system, which is obtained by discarding cyclic variables, is shown to yield the P\"oschl--Teller system \cite{pt}. In Section 4 we extend the analysis to the case of an extremal rotating black hole
in $d=2n$. The near horizon limit is constructed for a spacial case that all $n$ rotation parameters are equal. The associated conformal mechanics is analyzed and the Hamiltonian of a new $(2n-2)$--dimensional reduced integrable spherical mechanics is given. Section 5 contains similar results for
$d=2n+1$. In the concluding Section 6 we summarize our results and discuss possible further developments.

\vspace{0.5cm}

\noindent
{\bf 2.  $2d$ integrable model related to $4d$ Kerr black hole}\\

\vspace{0.3cm}

\noindent
{\it 2.1. Near horizon geometry of $4d$ extremal Kerr black hole}\\

In Boyer--Lindquist--type coordinates the Kerr solution of the Einstein equations reads\footnote{Throughout the paper we use the mostly minus signature convention for the metric tensor and put $c=1$, $G=1$.}
\bea\label{kerr}
&&
ds^2=dt^2-\frac{\rho^2}{\Delta} dr^2-\rho^2 d\theta^2-\frac{2Mr}{\rho^2} {\left(dt-a \sin^2{\theta} d\phi \right)}^2-(r^2+a^2) \sin^2{\theta} d\phi^2,
\nonumber\\[2pt]
&&
\Delta=r^2+a^2-2Mr, \qquad \rho^2=r^2+a^2 \cos^2{\theta},
\eea
where $M$ is the mass and $a$ is the rotation parameter.
The isometry group of (\ref{kerr}) includes the time translation and the rotation around the symmetry axis
\be\label{TP}
t'=t+\a, \qquad
\phi'=\phi+\b.
\ee

Zeros of $\Delta$ determine the inner and outer horizons which coalesce for the
extremal solution. Denoting the corresponding value of the radial coordinate by $r_0$ and assuming $a$ to be positive,
from the equations $\Delta (r_0)=0$, $\Delta' (r_0)=0$ one finds
\be
r_0=M=a.
\ee
In what follows we discuss the extremal solution only.

The construction of the near horizon Kerr geometry is tricky and it deserves to be discussed in detail.
A natural definition of the near horizon limit, which implies the redefinition of the radial coordinate
\be\label{nhl}
r \quad \rightarrow \quad r_0 + \epsilon r_0 r,
\ee
followed by $\epsilon \rightarrow 0$, yields a degenerate metric. Yet,
an important observation is that (\ref{nhl}) transforms the radial term
$\frac{\rho^2}{\Delta} dr^2$ into $r_0^2 (1+\cos^2{\theta})\frac{dr^2}{r^2}$, the last factor of which is a part of the $AdS_2$ metric
$r^2 dt^2-\frac{dr^2}{r^2}$. This prompts one to extract $r^2 dt^2$ from the fourth term in the metric (\ref{kerr}) by a proper redefinition of the temporal coordinate $t$ and the azimuthal angle $\phi$ \cite{bh}. To be more precise, one rewrites the term as
\be\label{SP}
\frac{\Delta}{\rho^2} {\left(dt-a \sin^2{\theta} d\phi \right)}^2-\frac{(r^2+a^2)}{\rho^2} {\left(dt-a \sin^2{\theta} d\phi \right)}^2,
\ee
and treats the two contributions separately. The first of them yields the desired expression $r_0^2 (1+\cos^2{\theta}) r^2 dt^2$ provided (\ref{nhl}) is extended by the new prescriptions
\be\label{nhl1}
t \quad \rightarrow \quad \frac{2 r_0 t}{\epsilon}, \qquad  \phi \quad \rightarrow \quad \phi+\frac{t}{\epsilon}.
\ee
The second term in (\ref{SP}) is then combined with the rest in (\ref{kerr}) involving $dt$ and $d\phi$ so as to form the expression
$-\frac{\sin^2{\theta}}{\rho^2} {(a dt-(r^2+a^2) d\phi)}^2$, which behaves well in the limit (\ref{nhl}), (\ref{nhl1}).
Combining all the pieces together, one gets the extremal Kerr throat solution \cite{bh}
\be\label{BH}
ds^2=(1+\cos^2{\theta}) \left( r^2 dt^2-\frac{dr^2}{r^2}-d\theta^2\right)-\frac{4 \sin^2{\theta}}{(1+\cos^2{\theta})} {\left(r dt+d\phi \right)}^2,
\ee
where the overall factor of $r_0^2$ has been discarded.\footnote{Note that in different coordinates this metric was derived earlier in \cite{zas}. Conformal symmetry of the solution was not discussed in \cite{zas}, however.}
It is amazing that the near horizon solution does not involve any physical parameters.

Near the throat the isometry group is enhanced.
In addition to (\ref{TP}) it includes the
dilatation
\be\label{Tp1}
t'=t+\g t, \qquad r'=r-\g r,
\ee
and the special conformal transformation
\be\label{Sp}
t'=t+(t^2+\frac{1}{r^2}) \s, \qquad r'=r-2 tr\s, \qquad \phi'=\phi-\frac{2}{r} \sigma.
\ee
Altogether they form $SO(2,1) \times U(1)$ group.

\vspace{0.5cm}

\noindent
{\it 2.2. $4d$ conformal mechanics and its integrable reductions}\\

The static gauge action functional for a massive particle propagating near the horizon of the $4d$ extremal Kerr black
hole reads
\be\label{start}
S=
-m \int d t \sqrt{(1+\cos^2{\theta})\left[r^2-\dot r^2/r^2-\dot\t^2 \right]-\frac{4 \sin^2{\t}}{(1+\cos^2{\theta})} {\left[r+\dot\phi\right]}^2},
\ee
where the dot denotes the derivative with respect to $t$.
We choose the Hamiltonian formalism to analyze the model. Introducing momenta $(p_r,p_\t,p_\phi)$ canonically
conjugate to the configuration space variables $(r,\t,\phi)$, one can readily derive the Hamiltonian and
the generators of special conformal transformations and dilatations from the Killing vector fields
(\ref{TP}), (\ref{Tp1}), (\ref{Sp}). The general formula which links a Killing vector field with components
$\xi^n(x)$ to the first integral $\xi^n(x) g_{nm}(x) \frac{d x^n}{d s}$ of the geodesic equations yields
\bea\label{hamc}
&&
H=r \left( \sqrt{m^2 (1+\cos^2{\theta})+{(r p_r)}^2 +p_\t^2+{\left(\frac{1+\cos^2{\t}}{2 \sin{\t}}\right)}^2 p_\phi^2} -p_\phi \right), \quad D=t H+r p_r,
\nonumber\\[2pt]
&&
K=\frac{1}{r} \left( \sqrt{m^2 (1+\cos^2{\theta})+{(r p_r)}^2 +p_\t^2+{\left(\frac{1+\cos^2{\t}}{2 \sin{\t}}\right)}^2 p_\phi^2} +p_\phi \right)+t^2 H+2trp_r.
\eea
Under the Poisson bracket the functions obey the structure relations of
$so(2,1)$ algebra
\be\label{confalg}
\{H,D \}=H, \quad \{H,K \}=2D, \quad \{D,K \} =K.
\ee
The Killing vector corresponding to the rotation symmetry leads to the conserved momentum $p_\phi$.

Like for a generic conformal mechanics \cite{ar,ar1}, the Casimir element of $so(2,1)$ realized in the model of a massive relativistic particle propagating near the horizon of an extremal black hole gives rise to a reduced Hamiltonian system \cite{gn}. For the case at hand one finds a two--dimensional system governed by the Hamiltonian
\be\label{HH1}
\mathcal{H}=p_\t^2+\left({\left[\frac{1+\cos^2{\t}}{2 \sin{\t}}\right]}^2 -1\right) p_\phi^2+m^2 (1+ \cos^2{\t}),
\ee
where $m$ is now treated as a coupling constant. As $p_\phi$ commutes with $\mathcal{H}$, the system is Liouville integrable.
For generic values of the parameters $m$ and $p_\phi$ a solution of the canonical equations of motion, which follow from (\ref{HH1}), involves elliptic integrals \cite{g2,bny}.

Because $\phi$  is cyclic, it is worth considering a further reduction of (\ref{HH1}) to a one--dimensional system, which is obtained by setting the conserved momentum $p_\phi$ to be a coupling constant directly in the Hamiltonain
\be\label{H2}
\mathcal{\tilde H}_{red}=p_\t^2+g^2 \cot^2{\t}+\nu \cos^2{\t}.
\ee
Here $g$ and $\nu$ are coupling constants related to the parameters of the original particle via  $g^2=p_\phi^2$ and $\nu=m^2-\frac 14 p_\phi^2$. As usual, the Hamiltonian is defined up to an additive constant. The resulting system (\ref{H2}) is a variant of the celebrated Higgs oscillator \cite{Higgs} coupled to an external field. 
\vspace{0.5cm}

\noindent
{\bf 3.  $3d$ integrable model related to $5d$ Myers--Perry black hole}\\

\vspace{0.3cm}

\noindent
{\it 3.1. Near horizon geometry of the extremal Myers--Perry black hole}\\

A generalization of the Kerr solution to the case of a five--dimensional spacetime was constructed by Myers and Perry \cite{mp}
\bea\label{mp}
&&
ds^2=dt^2-\frac{\rho^2}{\Delta} dr^2-\rho^2 d\theta^2-\frac{2M}{\rho^2} {\left(dt-a \sin^2{\theta} d\phi -b \cos^2{\theta} d\psi \right)}^2-
\nonumber\\[2pt]
&&
\quad \quad -(r^2+a^2) \sin^2{\theta} d\phi^2-(r^2+b^2) \cos^2{\theta} d\psi^2,
\nonumber\\[2pt]
&&
\Delta=\frac{1}{r^2} (r^2+a^2)(r^2+b^2)-2M, \quad \rho^2=r^2+a^2 \cos^2{\theta}+b^2 \sin^2{\theta},
\eea
where $M$ is the mass and $a$, $b$ are rotation parameters.
The isometry group of (\ref{mp}) is ${U(1)}^3$, which includes the time translation and two rotations
\be\label{TP1}
t'=t+\a, \qquad
\phi'=\phi+\b, \qquad \psi'=\psi+\gamma.
\ee
Note that, if the rotation parameters $a$ and $b$ are equal to each other, the isometry group of the Myers--Perry metric is enhanced
to $U(2)\times {U(1)}$ \cite{fs}. The extra symmetries read
\begin{align}\label{extra}
&
\t'=\t+\mu \cos{(\psi-\phi)}, && \t'=\t-\nu \sin{(\psi-\phi)},
\nonumber
\\[2pt]
&
\phi'=\phi+\mu \cot{\t} \sin{(\psi-\phi)}, && \phi'=\phi+\nu \cot{\t} \cos{(\psi-\phi)},
\nonumber
\\[2pt]
&
\psi'=\psi+\mu \tan{\t} \sin{(\psi-\phi)}, && \psi'=\psi+\nu \tan{\t} \cos{(\psi-\phi)},
\end{align}
where $\mu$ and $\nu$ are infinitesimal parameters.

The extremal solution occurs if $\Delta$ has a double zero at the horizon radius $r_0$.
Assuming $a$ and $b$ to be positive, from $\Delta (r_0)=0$ and $\Delta' (r_0)=0$ one finds
\be
r_0^2=ab ,\qquad M=\frac{{(a+b)}^2}{2}.
\ee
In what follows we will also need the relation
\be\label{Rel}
\lim_{r \rightarrow r_0} \frac{\Delta}{{(r-r_0)}^2}=4.
\ee

The near horizon limit of the Myers--Perry solution is constructed by analogy with the four dimensional case. One considers the second and the fourth terms in the metric (\ref{mp}) and rewrites them in the form
\be\label{mis}
-\frac{\rho^2}{\Delta} dr^2+\frac{\Delta}{\rho^2} {\left(dt-a \sin^2{\theta} d\phi -b \cos^2{\theta} d\psi \right)}^2-
\frac{(r^2+a^2)(r^2+b^2)}{r^2 \rho^2} {\left(dt-a \sin^2{\theta} d\phi -b \cos^2{\theta} d\psi \right)}^2.
\ee
The first two contributions entering (\ref{mis}) are reserved to produce the $AdS_2$ metric up to a factor. The last term in (\ref{mis}) and the rest in
(\ref{mp}) are combined together to form
\bea
&&
-\rho^2 d\t^2-\frac{\sin^2{\t}}{\rho^2} {[a dt-(r^2+a^2) d\phi]}^2-\frac{\cos^2{\t}}{\rho^2} {[b dt-(r^2+b^2) d\psi]}^2
\nonumber\\[2pt]
&&
-\frac{1}{r^2 \rho^2} {[a b dt-b(r^2+a^2) \sin^2{\t} d\phi -a(r^2+b^2) \cos^2{\t} d\psi]}^2.
\eea
At this stage one redefines the coordinates
\be
r \quad \rightarrow \quad r_0 + \epsilon r_0 r, \qquad t \quad \rightarrow \quad \frac{\alpha t}{\epsilon}, \qquad \phi \quad \rightarrow \quad \phi+\frac{\beta t}{\epsilon}, \qquad \psi \quad \rightarrow \quad \psi+\frac{\gamma t}{\epsilon},
\ee
and adjust the number coefficients $\alpha$, $\beta$, $\gamma$ so as to produce a finite result as $\epsilon \to 0$. Choosing
\be
\alpha=\frac{{(a+b)}^2}{4 r_0}, \qquad \beta=\gamma=\frac{(a+b)}{4 r_0},
\ee
and making use of (\ref{Rel}), one finally gets \cite{lmp}
\bea\label{nhm}
&&
ds^2=\rho_0^2 \left[r^2 dt^2-\frac{dr^2}{r^2}-4 d\t^2\right]-\frac{a b {(a+b)}^2 \sin^2{\t}}{\rho_0^2} {\left[r dt+d \phi\right]}^2
\\
&& \qquad
-\frac{a b {(a+b)}^2 \cos^2{\t}}{\rho_0^2} {\left[r dt+d \psi\right]}^2-\frac{1}{\rho_0^2} {\left[\rho_0^2 r dt+b(a+b)\sin^2{\t} d \phi+a(a+b)\cos^2{\t} d \psi\right]}^2,
\nonumber\\[2pt]
&&
\rho_0^2=ab+a^2 \cos^2{\t}+b^2 \sin^2{\t},
\nonumber
\eea
where for simplicity we discarded the overall factor of $1/4$ and scaled $\frac{2a}{r_0} \phi \rightarrow \phi$, $\frac{2b}{r_0} \psi \rightarrow \psi$.

The near horizon metric has a larger symmetry. In addition to (\ref{TP1}), the isometry group of (\ref{nhm}) includes the
dilatation
\be\label{tp1}
t'=t+\lambda t, \qquad r'=r-\lambda r,
\ee
and the special conformal transformation
\be\label{sp}
t'=t+(t^2+\frac{1}{r^2}) \s, \qquad r'=r-2 tr\s, \qquad
\phi'=\phi-\frac{2}{r} \s, \qquad
\psi'=\psi-\frac{2}{r} \s,
\ee
which altogether form $SO(2,1) \times U(1)^2$. For equal rotation parameters this is further extended to  $SO(2,1) \times U(2)$.
The extra symmetries can are realized as in (\ref{extra}), provided one scales the azimuthal coordinates entering the near horizon metric as follows:
$\phi \rightarrow 2 \phi$, $\psi \rightarrow 2\psi$.

\vspace{0.5cm}

\noindent
{\it 3.2. $5d$ conformal mechanics and its integrable reductions}\\

Given the near horizon metric (\ref{nhm}), the static gauge action functional of a massive particle propagating on this background reads
\bea\label{AC}
&&
S=-m \int d t \left[ \rho_0^2 \left(r^2 - \frac{{\dot r}^2}{r^2}-4 \dot\t^2\right) -
\frac{a b {(a+b)}^2}{\rho_0^2} \left( \sin^2{\t} {(r+\dot\phi)}^2+\cos^2{\t} {(r+\dot\psi)}^2 \right)
\right.
\nonumber\\[2pt]
&& \qquad \qquad
{\left.
-\frac{1}{\rho_0^2} {\left(r \rho_0^2 +b(a+b)\sin^2{\theta} \dot\phi+a(a+b)\cos^2{\theta} \dot\psi\right)}^2 \right]}^{1/2},
\eea
where $\rho_0^2=ab+a^2 \cos^2{\t}+b^2 \sin^2{\t}$ and $m$ is the particle mass.
Introducing momenta $(p_r,p_\t,p_\phi,p_\psi)$ canonically
conjugate to the configuration space variables $(r,\t,\phi,\psi)$ and taking into account the explicit form of the Killing vectors
specified in the preceding section, one derives the Hamiltonian and the
generators of dilatations and special conformal transformations
\bea\label{Ham}
&&
H=r \sqrt{m^2 \rho_0^2+{(r p_r)}^2+\frac 14 p_\t^2+
\frac{\rho_0^4}{a b {(a+b)}^2} \left( \frac{p_\phi^2}{\sin^2{\t}}+\frac{p_\psi^2}{\cos^2{\t}}
-\frac{1}{\rho_0^2} {(b p_\phi+a p_\psi)}^2 \right)}-
\nonumber\\[2pt]
&&
\qquad
-r (p_\phi+p_\psi),
\nonumber\\[2pt]
&&
K=
\frac{1}{r}\sqrt{m^2 \rho_0^2+{(r p_r)}^2+\frac 14 p_\t^2+
\frac{\rho_0^4}{a b {(a+b)}^2} \left( \frac{p_\phi^2}{\sin^2{\t}}+\frac{p_\psi^2}{\cos^2{\t}}
-\frac{1}{\rho_0^2} {(b p_\phi+a p_\psi)}^2 \right)}+
\nonumber\\[2pt]
&&
\qquad
+\frac{1}{r}(p_\phi+p_\psi) +t^2 H+2 t r p_r, \qquad D=t H+r p_r.
\eea
The Killing vectors corresponding to rotations reproduce the conserved momenta $p_\phi$ and $p_\psi$.

The Casimir element of $so(2,1)$ specifies the Hamiltonian of a reduced $3d$ system
\be\label{H1}
\mathcal{H}=\frac 14 p_\t^2+
\frac{\rho_0^4}{a b {(a+b)}^2} \left( \frac{p_\phi^2}{\sin^2{\t}}+\frac{p_\psi^2}{\cos^2{\t}}
-\frac{1}{\rho_0^2} {(b p_\phi+a p_\psi)}^2 \right)-{(p_\phi+p_\psi)}^{2}+m^2 \rho_0^2,
\ee
where $\rho_0^2=ab+a^2 \cos^2{\t}+b^2 \sin^2{\t}$.
As $p_\phi$ and $p_\psi$ commute with the Hamiltonian, the system is Liouville integrable.

If the rotation parameters $a$ and $b$ of the original black hole are equal to each other, the Hamiltonian (\ref{H1})
simplifies to
\be\label{H4}
\mathcal{\tilde H}=\frac 14 p_\t^2+\frac{p_\phi^2}{\sin^2{\t}}+\frac{p_\psi^2}{\cos^2{\t}}
-\frac 32 {(p_\phi+p_\psi)}^{2}.
\ee
In accord with our consideration above, the model possesses four integrals of motion, which are linked to $U(2)$ factor in the isometry group
\bea
&&
J_0=p_\psi+p_\phi, \qquad
J_1=p_\psi-p_\phi,
\nonumber\\[2pt]
&&
J_2=\frac 12 p_\t \cos{\left({\textstyle\frac 12 } (\psi-\phi) \right) }+\left(p_\phi \cot{\t}+p_\psi \tan{\t}\right) \sin{\left({\textstyle\frac 12 } (\psi-\phi) \right) },
\nonumber\\[2pt]
&&
J_3= \frac 12
p_\t \sin{\left({\textstyle\frac 12 } (\psi-\phi) \right) }-\left( p_\phi \cot{\t}+p_\psi \tan{\t} \right) \cos{\left({\textstyle\frac 12 } (\psi-\phi) \right)}.
\eea
In particular, under the Poisson bracket $J_1$, $J_2$ and $J_3$ obey the structure relations of $su(2)$ and commute with $J_0$. The Hamiltonian is proportional to the Casimir element of $su(2)$
\be
\mathcal{\tilde H}=J_i J_i-\frac 32 J_0^2.
\ee
As there are four functionally independent integrals of motion in a system with three degrees of freedom,  (\ref{H4}) determines a minimally superintegrable model.

Because the variables $\phi$ and $\psi$ are cyclic, one might be interested in a further reduction of (\ref{H4}) to a one--dimensional system. Setting in (\ref{H4}) the momenta $p_\phi$ and $p_\psi$ to be coupling constants, one gets
\be\label{dih}
\mathcal{\tilde H}_{red}=\frac 14 p_\t^2+\frac{\nu_1^2}{\sin^2{\t}}+\frac{\nu_2^2}{\cos^2{\t}}.
\ee
This is a variant of the celebrated P\"oschl--Teller model \cite{pt}.

\vspace{0.5cm}

\noindent
{\bf 4.  Extremal black holes in $d=2n$ and integrable models }\\

\vspace{0.3cm}

\noindent
{\it 4.1. Near horizon geometry of the extremal maximally symmetric $d=2n$ black hole}\\

\vspace{0.3cm}

A generalization of the Kerr solution of the Einstein equations to the case of even--dimensional space--time  was proposed in \cite{mp}.
In Boyer--Lindquist--type coordinates it reads
\bea\label{ed}
&&
ds^2=dt^2-\frac{U}{\Delta} dr^2-\frac{2 M}{U} {\left(dt -\sum_{i=1}^{n-1} a_i \mu_i^2 d \phi_i\right)}^2-\sum_{i=1}^{n}(r^2+ a_i^2) d \mu_i^2
-\sum_{i=1}^{n-1}(r^2+ a_i^2) \mu_i^2 d\phi_i^2,
\nonumber\\[2pt]
&&
\Delta=\frac{1}{r} \prod_{i=1}^{n-1} (r^2+a_i^2)-2M, \qquad U=r \sum_{i=1}^n \frac{\mu_i^2}{r^2+a_i^2} \prod_{j=1}^{n-1} (r^2+a_j^2),
\eea
where the latitudinal coordinates $\mu_i$ obey the constraint
\be
\sum_{i=1}^n \mu_i^2=1.
\ee
It is assumed that only $(n-1)$ independent rotation parameters are present so $a_n$ is set to zero in (\ref{ed}). The range of azimuthal coordinates $\phi_i$ is taken to be $[0,2\pi]$, $\mu_i$ lie in the interval $[0,1]$ for $i=1,\dots, n-1$, while $\mu_n \in [-1,1]$. The isometry group of (\ref{ed}) includes the time translation and $(n-1)$ rotations which altogether form $U(1)^n$.

Because in this work we are primarily concerned with the construction of superintegrable systems, in what follows we consider only the special case that all the rotation parameters are equal $a_i=a$. In this case the $U(1)^{n-1}$ subgroup in the isometry group, which corresponds to rotations, is known to enhance to $U(n-1)$ (see e.g. the discussion in \cite{vsp}). Integrable models which follow from less symmetric configurations will be studied elsewhere. Note that for unequal constants $a_i$ the near horizon metric has been constructed in \cite{lmp}. However, the case $a_i=a$ can not be derived from \cite{lmp} as the metric blows up for equal rotation parameters.

Before implementing the near horizon limit one has to put the metric in a convenient form.
Following the steps outlined in the preceding sections, one finds
\bea\label{ed1}
&&
ds^2=\frac{\Delta}{U} {\left(dt -a \sum_{i=1}^{n-1} \mu_i^2 d \phi_i\right)}^2-\frac{U}{\Delta} dr^2
-\frac{{(r^2+a^2)}^{n-2}}{r U} \sum_{i=1}^{n-1} \mu_i^2 {\left(a dt-(r^2+a^2) d\phi_i\right)}^2
\nonumber\\[2pt]
&& \qquad-(r^2+a^2) \sum_{i=1}^{n-1} d\mu_i^2-r^2 d \mu_n^2 +\frac{a^2 {(r^2+a^2)}^{n-1}}{r U} \sum_{i<j}^{n-1} \mu_i^2 \mu_j^2 {\left(d\phi_i-d\phi_j \right)}^2,
\\[2pt]
&&
\Delta=\frac{1}{r} {(r^2+a^2)}^{n-1}-2M, \quad U=\frac{1}{r} {(r^2+a^2)}^{n-2} (r^2+a^2 \mu_n^2), \quad \mu_n^2=1-\sum_{i=1}^{n-1} \mu_i^2.
\nonumber
\eea
The extremal solution is characterized by the condition that $\Delta$ has a double zero at the horizon radius $r=r_0$. The conditions $\Delta(r_0)=0$ and $\Delta'(r_0)=0$ allow one to fix the mass $M$ and the rotation parameter $a$ in terms of $r_0$
\be
M=\frac{r_0^{2n-3} {[2(n-1)]}^{n-1}}{2}, \qquad a^2=(2n-3) r_0^2.
\ee
In what follows we will also need the relation
\be\label{supcond}
\lim_{r \to r_0} \frac{\Delta}{{(r-r_0)}^2}=\frac{(2n-3){[2(n-1) r_0^2 ]}^{n-2}}{r_0}.
\ee

In order to implement the near horizon limit, one redefines the coordinates
\be
r \quad \rightarrow \quad r_0 + \epsilon r_0 r, \qquad t \quad \rightarrow \quad \frac{\alpha t}{\epsilon}, \qquad \phi_i \quad \rightarrow \quad \phi_i+\frac{\beta_i t}{\epsilon},
\ee
and adjusts the number coefficients $\alpha$ and $\beta_i$ in such a way that, up to a factor, the first two terms in (\ref{ed1}) produce the $AdS_2$ metric in the limit $\epsilon \to 0$
\be
\alpha=\frac{2(n-1) r_0}{2n-3}, \qquad \beta_i=\frac{r_0}{a}.
\ee
Sending $\epsilon$ to zero and taking into account (\ref{supcond}), one derives the near horizon metric
\bea\label{Nhm}
&&
ds^2=\rho_0^2 \left(r^2 dt^2-\frac{dr^2}{r^2} \right)-2(n-1)\sum_{i=1}^{n-1} d\mu_i^2-d\mu_n^2-\frac{4}{{(2n-3)}^2 \rho_0^2} \sum_{i=1}^{n-1} \mu_i^2 {(r dt+d \phi_i)}^2+
\nonumber\\[2pt]
&&
\qquad \quad +\frac{2}{(n-1)(2n-3) \rho_0^2} \sum_{i<j}^{n-1} \mu_i^2 \mu_j^2 {(d \phi_i-d\phi_j)}^2,
\\[2pt]
&&
\rho_0^2=\frac{1+(2n-3) \mu_n^2}{2n-3}, \qquad \mu_n^2=1-\sum_{i=1}^{n-1} \mu_i^2,
\nonumber
\eea
where we discarded an overall factor of $r_0^2$ and scaled the azimuthal angular variables as follows: $\frac{a(n-1)}{r_0} \phi_i \rightarrow \phi_i$. It is straightforward to verify that (\ref{Nhm}) is a vacuum solution of the Einstein equations.
Note that setting $n=2$, $\mu_1=\sin{\t}$, $\mu_2=\cos{\t}$ (with $\t\in[0,\pi]$) one reproduces the Bardeen--Horowitz metric (\ref{BH}).

Similarly to the lower dimensional patterns considered above, (\ref{Nhm}) exhibits conformal symmetry, which
is realized as in Eqs. (\ref{Tp1}) and (\ref{Sp}) with the obvious alteration of the special conformal transformation 
\be\label{sp}
\phi'_i=\phi_i-\frac{2}{r} \s,
\ee
acting on the azimuthal angular variables.

\vspace{0.5cm}

\noindent
{\it 4.2. Conformal mechanics in $d=2n$ and its integrable reductions}\\

\vspace{0.3cm}
The construction of the Hamiltonian of a massive relativistic particle propagating on the curved background (\ref{Nhm}) is a straightforward,
albeit somewhat tedious task. An alternative method is to invert the metric (\ref{Nhm})
\bea\label{inv}
&&
g^{mn}(x) \partial_n \partial_m=\frac{1}{r^2 \rho_0^2}  {\left(\frac{\partial}{\partial t}\right)}^2-\frac{r^2}{\rho_0^2}  {\left(\frac{\partial}{\partial r}\right)}^2-\frac{2}{r \rho_0^2} \sum_{i=1}^{n-1} \frac{\partial}{\partial t} \frac{\partial}{\partial \phi_i}+
\nonumber\\[2pt]
&&
\qquad \quad +\frac{1}{(2n-3) (2n-2) \rho_0^2} \sum_{i,j=1}^{n-1}(\mu_i \mu_j-(2n-3) \rho_0^2 \delta_{ij}) \frac{\partial}{\partial \mu_i} \frac{\partial}{\partial \mu_j}+
\nonumber\\[2pt]
&&
\qquad \quad +\sum_{i,j=1}^{n-1}\left(\frac{{(2n-3)}^2}{4}+\frac{1}{\rho_0^2} -\frac{(2n-3)(2n-2)}{4 \mu_i^2} \delta_{ij}
\right) \frac{\partial}{\partial \phi_i} \frac{\partial}{\partial \phi_j},
\nonumber\\[2pt]
&&
\rho_0^2=\frac{2(n-1)}{2n-3}-\sum_{i=1}^{n-1} \mu_i^2,
\eea
where $\delta_{ij}$ is the Kronecker delta, and then solve the mass shell condition
\be
g^{nm} p_n p_m=m^2,
\ee
where $p_m=(p_0,p_r,p_{\mu_i},p_{\phi_i})$, for $p_0$
\bea\label{p0}
&&
p_0=H=r \left( \left[m^2 \rho_0^2+{(r p_r)}^2-\frac{1}{(2n-3)(2n-2)} \sum_{i,j=1}^{n-1}(\mu_i \mu_j-(2n-3) \rho_0^2 \delta_{ij}) p_{\mu_i} p_{\mu_j} \right. \right.
\nonumber\\[2pt]
&&
\left.{\left.-\sum_{i,j=1}^{n-1}\left(1+\frac{{(2n-3)}^2 \rho_0^2}{4}-\frac{(2n-3)(2n-2)\rho_0^2}{4 \mu_i^2} \delta_{ij}
\right) p_{\phi_i} p_{\phi_j}+{\left(\sum_{i=1}^{n-1} p_{\phi_i}\right)}^2 \right]}^{1/2} -\sum_{i=1}^{n-1} p_{\phi_i} \right),
\nonumber\\[2pt]
&&
\eea
where we redefined the momenta $p_{\phi_i} \to -p_{\phi_i}$ so as to conform with the notation adopted for the Hamiltonian mechanics in Sec. 2.2.
Eq. (\ref{p0}) gives the Hamiltonian of a conformal mechanics in $d=2n$. Conformal generators are constructed by analogy with the lower dimensional patterns considered above
\bea
&&
K=\frac{1}{r} \left( \left[m^2 \rho_0^2+{(r p_r)}^2-\frac{1}{(2n-3)(2n-2)} \sum_{i,j=1}^{n-1}(\mu_i \mu_j-(2n-3) \rho_0^2 \delta_{ij}) p_{\mu_i} p_{\mu_j}- \right. \right.
\nonumber\\[2pt]
&&
\left.{\left.-\sum_{i,j=1}^{n-1}\left(1+\frac{{(2n-3)}^2 \rho_0^2}{4}-\frac{(2n-3)(2n-2)\rho_0^2}{4 \mu_i^2} \delta_{ij}
\right) p_{\phi_i} p_{\phi_j}+{\left(\sum_{i=1}^{n-1} p_{\phi_i}\right)}^2 \right]}^{1/2} +\sum_{i=1}^{n-1} p_{\phi_i} \right)
\nonumber\\[2pt]
&&
\quad \quad
+t^2 H+2trp_r, \qquad D=t H+r p_r.
\eea
Note that setting $n=2$, $\mu_1=\sin{\t}$, $p_{\mu_1}=\frac{p_{\t}}{\cos{\t}}$, with $(\t,p_\t)$ being a canonical pair, one
reproduces the Hamiltonian of the conformal mechanics in four dimensions (\ref{hamc}).

Computing the Casimir element of $so(2,1)$, one gets the Hamiltonian of a reduced integrable spherical mechanics
\bea\label{Heven}
&&
\mathcal{\tilde H}=\frac{1}{(2n-3)(2n-2)} \sum_{i,j=1}^{n-1}((2n-3) \rho_0^2 \delta_{ij}-\mu_i \mu_j) p_{\mu_i} p_{\mu_j}+
\nonumber\\[2pt]
&& \quad \quad
+\sum_{i,j=1}^{n-1}\left(\frac{(2n-3)(2n-2)\rho_0^2}{4 \mu_i^2} \delta_{ij}-\frac{{(2n-3)}^2 \rho_0^2}{4}-1
\right) p_{\phi_i} p_{\phi_j}+m^2 \rho_0^2,
\eea
where $\rho_0^2$ is given in (\ref{inv}) and $m^2$ is now treated as a coupling constant. By construction, it inherits $U(n-1)$ symmetry of the background. A further reduction occurs if one disregards the cyclic variables $\phi_i$ and sets $p_{\phi_i}$  to be constants in (\ref{Heven})
\bea\label{final}
&&
{\mathcal{\tilde H}}_{red}=\frac{1}{(2n-3)(2n-2)} \sum_{i,j=1}^{n-1}((2n-3) \rho_0^2 \delta_{ij}-\mu_i \mu_j) p_{\mu_i} p_{\mu_j}+\sum_{i=1}^{n-1} \frac{g_i^2 \rho_0^2}{\mu_i^2}+\nu \sum_{i=1}^{n-1} \mu_i^2,
\nonumber\\[2pt]
&&
\rho_0^2=\frac{2(n-1)}{2n-3}-\sum_{i=1}^{n-1} \mu_i^2.
\eea
Here $\nu$ and $g_i$ are coupling constants. Thus, we have constructed a new variant of integrable spherical mechanics with $2(n-1)$ phase space degrees of freedom.

\vspace{0.5cm}

\noindent
{\bf 5.  Extremal black holes in $d=2n+1$ and integrable models }\\

\vspace{0.3cm}

\noindent
{\it 5.1. Near horizon geometry of the extremal maximally symmetric $d=2n+1$ black hole}\\

\vspace{0.3cm}

If the dimension of spacetime is odd, there are $n$ azimuthal angular coordinates $\phi_i$ and respectively $n$ rotation parameters $a_i$.
A vacuum solution of the Einstein equations describing a black hole in $2n+1$ dimensions, which rotates in $n$ orthogonal spatial $2$-planes, reads
\cite{mp}
\bea\label{odd}
&&
ds^2=dt^2-\frac{U}{\Delta} dr^2-\frac{2 M}{U} {\left(dt -\sum_{i=1}^{n} a_i \mu_i^2 d \phi_i\right)}^2-\sum_{i=1}^{n}(r^2+ a_i^2) (d \mu_i^2
+\mu_i^2 d\phi_i^2),
\nonumber\\[2pt]
&&
\Delta=\frac{1}{r^2} \prod_{i=1}^{n} (r^2+a_i^2)-2M, \qquad U=\sum_{i=1}^n \frac{\mu_i^2}{r^2+a_i^2} \prod_{j=1}^{n} (r^2+a_j^2).
\eea
As before, the latitudinal coordinates $\mu_i$ are subject to the constraint
\be\label{lc}
\sum_{i=1}^n \mu_i^2=1,
\ee
and it is assumed that all $\mu_i$ lie in the interval $[0,1]$. The isometry group of (\ref{odd}) includes the time translation and $n$ rotations acting on the azimuthal variables.

In what follows we focus on the extremal and maximally symmetric configuration, for which
\be
a_i=a, \qquad   M=\frac{n^n {r_0}^{2n-2}}{2}, \qquad a^2=(n-1) r_0^2.
\ee
The last two conditions follow from the requirement that $\Delta(r)$ has a double zero at the horizon radius $r=r_0$.
Note that in this case the rotational symmetry group is known to enhance from ${U(1)}^n$ to $U(n)$.

In order to construct the near horizon metric, we rewrite (\ref{odd}) (where all the rotation parameters are set equal) in the form
\bea\label{ed2}
&&
ds^2=\frac{\Delta}{U} {\left(dt -a \sum_{i=1}^{n} \mu_i^2 d \phi_i\right)}^2-\frac{U}{\Delta} dr^2
-\frac{1}{r^2} \sum_{i=1}^{n} \mu_i^2 {\left(a dt-(r^2+a^2) d\phi_i\right)}^2
\nonumber\\[2pt]
&& \qquad-(r^2+a^2) \sum_{i=1}^{n} d \mu_i^2+\frac{a^2 (r^2+a^2)}{r^2} \sum_{i<j}^{n} \mu_i^2 \mu_j^2 {\left(d\phi_i-d\phi_j \right)}^2,
\\[2pt]
&&
\Delta=\frac{{(r^2+a^2)}^n}{r^2}-2M, \quad U={(r^2+a^2)}^{n-1}, \quad \mu_n^2=1-\sum_{i=1}^{n-1} \mu_i^2,
\nonumber
\eea
and redefine the coordinates
\be
r \quad \rightarrow \quad r_0 + \epsilon r_0 r, \qquad t \quad \rightarrow \quad \frac{n r_0 t}{2(n-1)\epsilon}, \qquad \phi_i \quad \rightarrow \quad \phi_i+\frac{r_0 t}{2 a \epsilon}.
\ee
Then we
take into account the relation
\be
\lim_{r \to r_0} \frac{\Delta(r)}{{(r-r_0)}^2}=\frac{2(n-1) {(n r_0^2)}^{n-1}}{r_0^2},
\ee
and send $\epsilon$ to zero. This yields
\bea\label{Nhm1}
&&
ds^2=r^2 dt^2-\frac{dr^2}{r^2}-2n(n-1)\sum_{i=1}^{n} d\mu_i^2-2 \sum_{i=1}^{n} \mu_i^2 {(r dt+d \phi_i)}^2+
\nonumber\\[2pt]
&&
\qquad \quad +\frac{2(n-1)}{n} \sum_{i<j}^{n} \mu_i^2 \mu_j^2 {(d \phi_i-d\phi_j)}^2, \qquad \mu_n^2=1-\sum_{i=1}^{n-1} \mu_i^2,
\eea
where we discarded an overall factor of $\frac{r_0^2}{2(n-1)}$ and scaled the azimuthal variables as follows: $\frac{an}{r_0} \phi_i \rightarrow \phi_i$. It is straightforward to verify that (\ref{Nhm1}) is a vacuum solution of the Einstein equations.
Note that setting $n=2$, $\mu_1=\sin{\t}$ and $\mu_2=\cos{\t}$, one reproduces the metric (\ref{nhm}) with $a=b$ and an overall factor of $\rho_0^2=2a^2$ being discarded.

\vspace{0.5cm}

\noindent
{\it 5.2. Conformal mechanics in $d=2n+1$ and its integrable reductions}\\

\vspace{0.3cm}
In order to construct the Hamiltonian of a conformal mechanics, which is associated with the background geometry (\ref{Nhm1}), we invert the metric
\bea\label{inv1}
&&
g^{mn}(x) \partial_n \partial_m=\frac{1}{r^2}  {\left(\frac{\partial}{\partial t}\right)}^2-r^2 {\left(\frac{\partial}{\partial r}\right)}^2-\frac{2}{r} \sum_{i=1}^{n} \frac{\partial}{\partial t} \frac{\partial}{\partial \phi_i}+\frac{1}{2n(n-1)} \sum_{i,j=1}^{n-1}(\mu_i \mu_j-\delta_{ij}) \frac{\partial}{\partial \mu_i} \frac{\partial}{\partial \mu_j}
\nonumber\\[2pt]
&&
\qquad \qquad \qquad +\sum_{i,j=1}^{n}\left(\frac{(n+1)}{2}-\frac{n}{2 \mu_i^2} \delta_{ij}
\right) \frac{\partial}{\partial \phi_i} \frac{\partial}{\partial \phi_j},
\qquad
\mu_n^2=1-\sum_{i=1}^{n-1} \mu_i^2,
\eea
where $\delta_{ij}$ is the Kronecker delta, and solve the mass shell condition $g^{nm} p_n p_m=m^2$
for $p_0$\footnote{In (\ref{P0}) the momenta $p_{\phi_i}$ were redefined $p_{\phi_i} \to -p_{\phi_i}$ so as to conform to the notation adopted for the Hamiltonian mechanics in Sec. 3.2.}
\bea\label{P0}
&&
p_0=H=r \left( \left[m^2+{(r p_r)}^2+\frac{1}{2n(n-1)} \sum_{i,j=1}^{n-1}(\delta_{ij}-\mu_i \mu_j) p_{\mu_i} p_{\mu_j}
\right. \right.
\nonumber\\[2pt]
&& \qquad \qquad
\left.{\left.+\sum_{i,j=1}^{n}\left(\frac{n}{2 \mu_i^2} \delta_{ij}-\frac{(n+1)}{2}
\right) p_{\phi_i} p_{\phi_j}+{\left(\sum_{i=1}^{n} p_{\phi_i}\right)}^2 \right]}^{1/2} -\sum_{i=1}^{n} p_{\phi_i} \right).
\eea
The generators of special conformal transformations and dilatations read
\bea
&&
K=\frac{1}{r} \left( \left[m^2+{(r p_r)}^2+\frac{1}{2n(n-1)} \sum_{i,j=1}^{n-1}(\delta_{ij}-\mu_i \mu_j) p_{\mu_i} p_{\mu_j}+
\right. \right.
\nonumber\\[2pt]
&& \qquad \qquad
\left.{\left.+\sum_{i,j=1}^{n}\left(\frac{n}{2 \mu_i^2} \delta_{ij}-\frac{(n+1)}{2}
\right) p_{\phi_i} p_{\phi_j}+{\left(\sum_{i=1}^{n} p_{\phi_i}\right)}^2 \right]}^{1/2} +\sum_{i=1}^{n} p_{\phi_i} \right)+
\nonumber\\[2pt]
&&
\qquad \quad
+t^2 H+2trp_r, \qquad D=t H+r p_r.
\eea

As usual, the Hamiltonian of a reduced integrable spherical mechanics is determined by the Casimir element of $so(2,1)$
\bea\label{Hodd}
&&
\mathcal{\tilde H}=\frac{1}{2n(n-1)} \sum_{i,j=1}^{n-1}(\delta_{ij}-\mu_i \mu_j) p_{\mu_i} p_{\mu_j}+
\sum_{i,j=1}^{n}\left(\frac{n}{2 \mu_i^2} \delta_{ij}-\frac{(n+1)}{2}
\right) p_{\phi_i} p_{\phi_j}.
\eea
By construction, it inherits $U(n)$ symmetry of the background. Ignoring the cyclic variables $\phi_i$, one obtains a further reduction
\bea\label{Hodd1}
&&
{\mathcal{\tilde H}}_{red}=\frac{1}{2n(n-1)} \sum_{i,j=1}^{n-1}(\delta_{ij}-\mu_i \mu_j) p_{\mu_i} p_{\mu_j}+
\sum_{i=1}^{n}\frac{g_i^2}{\mu_i^2}, \qquad \mu_n^2=1-\sum_{i=1}^{n-1} \mu_i^2,
\eea
where $g_i^2$ are coupling constants. This new integrable system can be interpreted as a particle moving on $(n-1)$--dimensional sphere and interacting with an external field. Setting $n=2$, $\mu_1=\sin{\t}$ and $p_{\mu_1}=\frac{p_{\t}}{\cos{\t}}$, with $(\t,p_\t)$ being a canonical pair, one
reproduces the P\"oschl--Teller Hamiltonian (\ref{dih}).

\vspace{0.5cm}

\noindent
{\bf 6. Conclusion}

\vspace{0.3cm}
To summarize, in this work we have constructed a metric describing the near horizon geometry of
an extremal rotating black hole in arbitrary dimension for the special case that all the rotation parameters are equal.
The Hamiltonian of an integrable spherical mechanics associated with such a geometry was derived.

Many issues related to the present work deserve a further study. First of all, it is important to explicitly describe $U(n)$ symmetry underlying the models beyond five dimensions. This will require an explicit parametrization of the sphere in the sector of latitudinal variables.
As was demonstrated in Section 2 and Section 3, the most interesting models
arise if one considers a reduction of the spherical mechanics, which is obtained by discarding cyclic variables.
It is interesting to describe these models in full generality and to uncover their superintegrability.
In odd--dimensional space--time such a system describes a particle moving on $(n-1)$--dimensional sphere and interacting with an external field. Geometry behind a similar model (\ref{final}) corresponding to an even--dimensional space--time is of considerable interest.
A possible relation between the models constructed in this work and $A_n$--models in the classification of Olshanetsky and Perelomov \cite{op} is worth studying.

In this work we have considered a maximally symmetric black hole configuration, which in general yields a superintegrable spherical mechanics.
Less symmetric configurations, for which not all the rotation parameters are equal, will result in a spherical mechanics with less symmetry, which, however, may maintain the property of integrability. It would be interesting to derive a constraint on the rotation parameters of a black hole, which
guarantees that the resulting spherical mechanics is integrable. Another interesting issue is a generalization of the present analysis to the case of
the near horizon extremal rotating black hole on AdS background. A nonvanishing cosmological constant will yield a more complicated potential for the
spherical mechanics.

\vspace{0.5cm}

\noindent{\bf Acknowledgements}\\

\noindent
We thank Olaf Lechtenfeld and Armen Nersessian for their interest in the work and reading an earlier version of the manuscript. This work was supported by RF Federal Program "Kadry" under the contracts 16.740.11.0469, 14.B37.21.0786
and MSE Program "Nauka" under the grant 1.604.2011.

\end{document}